\documentclass[showpacs,amsmath,amssymb,onecolumn,pra,superscriptaddress]{revtex4-1}
\usepackage{amssymb}
\usepackage[dvips]{graphicx}
\usepackage{enumerate}
\usepackage{epsfig}
\usepackage{subfigure}
\usepackage{xcolor}
\usepackage[T1]{fontenc}
\usepackage[expert]{mathdesign}
\usepackage{fullpage}
\usepackage{amsthm,amsfonts,amssymb,amscd,mathrsfs,eufrak,xspace,framed}
\usepackage{amsmath}
\usepackage{color}
\usepackage{setspace}
\usepackage{url}
\usepackage{enumitem}
\usepackage{soul}
\begin{document}

\title{Practical issues of twin-field quantum key distribution}

\author{Feng-Yu Lu}
\author{Zhen-Qiang Yin}
\email{yinzq@ustc.edu.cn}
\author{Rong-Wang}
\author{Guan-Jie Fan-Yuan}
\author{Shuang Wang}
\author{De-Yong He}
\author{Wei Chen}
\affiliation{CAS Key Laboratory of Quantum Information, CAS Center For Excellence in Quantum Information and Quantum Physics,
	University of Science and Technology of China, Hefei 230026, China}
\affiliation{State Key Laboratory of Cryptology, P. O. Box 5159, Beijing 100878, China}
\author{Wei Huang}
\author{Bing-Jie Xu}
\affiliation{Science and Technology on Communication Security Laboratory, Institute of Southwestern Communication, Chengdu, Sichuan 610041, China}
\author{Guang-Can Guo}
\author{Zheng-Fu Han}
\affiliation{CAS Key Laboratory of Quantum Information, CAS Center For Excellence in Quantum Information and Quantum Physics,
University of Science and Technology of China, Hefei 230026, China}
\affiliation{State Key Laboratory of Cryptology, P. O. Box 5159, Beijing 100878, China}

\begin{abstract}

Twin-Field Quantum Key Distribution(TF-QKD) protocol and its variants, such as Phase-Matching QKD(PM-QKD), sending or not QKD(SNS-QKD) and No Phase Post-Selection TF-QKD(NPP-TFQKD), are very promising for long-distance applications. However, there are still some gaps between theory and practice in these protocols. Concretely, a finite-key size analysis is still missing, and the intensity fluctuations are not taken into account. To address the finite-key size effect, we first give the key rate of NPP-TFQKD against collective attack in finite-key size region and then prove it can be against coherent attack. To deal with the intensity fluctuations, we present an analytical formula of 4-intensity decoy state NPP-TFQKD and a practical intensity fluctuation model. Finally, through detailed simulations, we show NPP-TFQKD can still keep its superiority of high key rate and long achievable distance.
\end{abstract}

\maketitle

\section{Introduction}
Quantum Key Distribution(QKD)\cite{bennett2014quantum,ekert1991quantum} is one of the most mature applications among the emerging quantum technologies. It allows two remote users, called Alice and Bob, to share random secret keys even if there is an eavesdropper, Eve\cite{mayers2001unconditional,lo1999unconditional,shor2000simple}. Due to the loss of channel, both the key rate and achievable distance of QKD are limited. Although increasing the secret key rate(SKR) and achievable distance are essentially significant for the real applications of QKD, the theorists proved there are some limits on the improvement of SKR\cite{takeoka2014fundamental,pirandola2017fundamental}. In particular, for the channel of transmittance $\eta$, the linear bound \cite{pirandola2017fundamental}, i.e. $R\leqslant -log_2(1-\eta)$, gives the precise SKR bound for any point-to-point QKD without quantum repeaters. Surprisingly, a revolutionary protocol called Twin-Field Quantum Key Distribution(TF-QKD)\cite{tfqkd} was recently proposed to beat this bound. Inspired by the novel idea of TF-QKD, researchers proposed some variants and completed the corresponding security proofs
\cite{Lo-tfqkd,pmqkd,sns-tfqkd,cui2018phase,curty2018simple,lin2018simple}. From the view of experiments, these variants, i.e. Phase-Matching QKD(PM-QKD)\cite{pmqkd}, sending or not QKD(SNS-QKD)\cite{sns-tfqkd} and No Phase Post-Selection TF-QKD(NPP-TFQKD)\cite{cui2018phase,curty2018simple,lin2018simple}, are simpler. Indeed, both the SNS-QKD and NPP-TFQKD have been scuccessfully demonstrated\cite{minder2019experimental,wang2019beating,liu2019experimental,zhong2019proof}.

However, there are still some gaps between theory and implementation of TF-QKD.
The first problem is the finite-key size effect is still not considered previously. In Refs.\cite{cui2018phase,curty2018simple,lin2018simple}, asymptotic SKR of NPP-TFQKD is proposed, but the SKR in finite-key region is not given. On the other hand, the key-size in a practical implementation is always finite, thus a framework to deal with the finite-key size effect in TF-QKD is indispensable.

Another problem we will discuss is a potential security loophole of TF-QKD and its variants.
Although the Refs.\cite{tfqkd,Lo-tfqkd,pmqkd,sns-tfqkd,cui2018phase,curty2018simple,lin2018simple} have proved 
the TF-QKD and its variants are information-theoretically secure even with unstrusted measurement device just like the original measurement-device-independent protocol\cite{lo2012measurement,wang2015phase,liu2013experimental,ma2012alternative}, the imperfections of laser source may spoil the security. One of the intractable loopholes of source is the intensity fluction\cite{wang2009decoy,wang2007decoy,zhou2017implementing}. In the existing security proofs of NPP-TFQKD, it is assumed that Alice and Bob are able to accurately control the intensity of signal and decoy modes, which is not perfectly satisfied in experiment. In this work, we also propose a countermeasure to tackle the internsity fluctuation of NPP-TFQKD. A key step of our method is proposing the analytical formulas to deal with the 4-intensity decoy states in NPP-TFQKD. In the original NPP-TFQKD\cite{cui2018phase}, one must use linear programing to solve linear equations of decoy states \cite{lo2005decoy,wang2005decoy,wang2005beating,wang2013three,yu2015statistical}. Compared with linear programing, analytical formula has superiorities on some special situations. More importantly, the proposed analytical formulas are particularly convinient to be incorporated to our intensity fluctuation. Another key step of our method is introducing a new intensity fluctuation model in finite-key size regime. The model makes TF-QKD robuster to intensity fluctuation.

The rest of this paper is organized as followoing. Firstly, In Sec.\textrm{II}, we briefly review the flow of NPP-TFQKD protocol. In Sec.\textrm{III}, we analyze the finite-key size effect of NPP-TFQKD, give the SKR formula against coherent attack and evaluate the performance of TF-QKD in finite-key regime. In Sec.\textrm{IV}, the analytical formulas for 4-intensity decoy state method are given. Then we introduce the intensity fluctuation model and its countermeasure. Finally, a completed simulation taken both the finite-key size effect and intensity fluctuation into account is present.

\section{Protocol definition}
\begin{figure}[htbp]
\includegraphics[width=9cm]{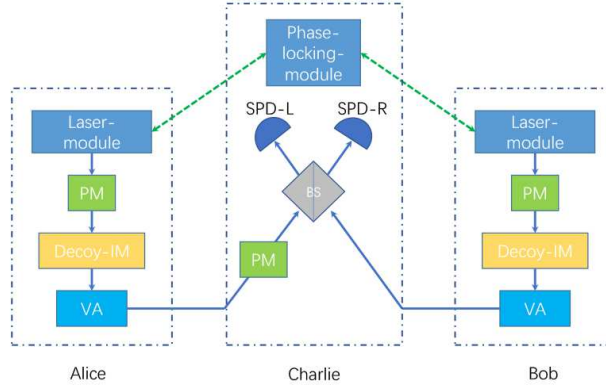}
\caption{\label{Fig_setup} In Alice and Bob's side, The laser-modules can prepare initial phase locked weak coherent pulses with the help of the phase-locking module belongs to Charlie\cite{wang2019beating}. The phase-modulators(PM) are apply to encode key bits in code mode and randomize the phase of WCPs in decoy mode. The intensity-modulators(decoy-IM) are set to modualte different intensities to apply decoy state technique. VA denotes variable attenuator. The PM belongs to Charlie can compensate phase drift caused by long-distance fiber. The gray square denotes 50:50 beam splitter and the blue semicircles are single photon detectors(SPDs)}
\end{figure}

  The setup of NPP-TFQKD\cite{cui2018phase} protocol is illustrated in Fig.\ref{Fig_setup} and the flow is as following:

\textbf{State preparation:} This step will be repeated by $N$ trials. In each trial, Alice(Bob) chooses code mode or decoy mode with probabilities $P_c$ and $P_d=1-P_c$ respectively, sends corresponding quantum state to untrusted Charlie.

When code mode is selected, Alice(Bob) prepares a phase-locked weak coherent pulse(WCP) $|\pm\sqrt{\mu}\rangle_{A}$($|\pm\sqrt{\mu}\rangle_{B}$), where the plus or minus of the quantum state depends on the bit value of Alice(Bob)'s random key of this trial.

When decoy mode is selected, Alice(Bob) prepares a phase randomized WCP, whose intensity ${\nu}_a$(${\nu}_b$) is randomly choosen from  a pre-decided set. Alice(Bob) actually prepares a mixed state since the randomized phase in the decay mode will never be publicly announced. For instance, the density matrix of Alice's WCP in decoy mode can be denoted as:
\begin{equation}
\rho_{\nu_a} = \sum_{n=0}^{\infty}{e^{-\nu_a}\frac{\nu_a^n}{n!}|n\rangle\langle{n}|},
\label{density_mat}
\end{equation}
where $|n\rangle$ is the Fock state.

\textbf{Measurement:} For each trail of the state preparation step, the untrusted Charlie must publicly announce a single click of his single photon detector(SPD) 'SPD-L' or 'SPD-R' or non-click meassage. Note that Charlie is untrusted, thus he is not necessarily to make the measurement shown in Fig.\ref{Fig_setup}.

\textbf{Sifting:} Alice and Bob publicly announce which trails are code mode and which are decoy mode. For the trials they both choose code mode and Charlie announce 'SPD-L' or 'SPD-R' clicked, Alice and Bob will retain this key bit. According to Charlie's measurement result, Bob may decide to flip his key bit or not. After this step, Alice and Bob generate sifted key bit string $Z$ and $Z'$ respectively.

\textbf{Error correction:} Alice sends $\lambda_{EC}=nfH(E_c)$ bits of classical error correction data to Bob. Here $n$=\big|Z\big|= \big|Z'\big| is the size of sifted key bits, $H(p) = -plog_2p - (1-p)log_2(1-p)$ is the Shannon entropy, $E_c$ is the error rate of sifted key bits and $f\geqslant 1$ denotes error correction efficiency. Depending on the error correction data and $Z'$, Bob obtains an estimated $\hat{Z}$ of $Z$. Next, by applying universal$_2$ hash fuction, Alice sends $\lambda_{EV}=log_2\frac{1}{\epsilon_{cor}}$ bits of error verification information to Bob. If the error verification fails, they output an empty string and abort the protocol. Otherwise, they assume the error correction sucesses and $Z=Z'$.

\textbf{Parameter estimation and privacy amplification:} Alice and Bob accumulate data to estimate gain $Q_c$ of trials that they both choose code mode, gains $Q_{xy}$ of trials they choose decoy mode with intensity $x$ and $y$ respectively. With these parameters and $\lambda_{EC}$, $\lambda_{EV}$, Alice and Bob perform privacy amplification, say, apply a random universal$_2$ hash function to $Z$ and $\hat{Z}$ respectively to generate $l_{sec}$-length secure bit string $S$ and $S'$ respectively. The SKR per pulse is defined as $R = l_{sec}/N$

\section{finite-key analysis of NPP-TFQKD}

Previous works\cite{cui2018phase,curty2018simple,lin2018simple} of 
NPP-TFQKD are based on the asymptotic situation. However, since it's impossible for Alice and Bob to send infinite pulses to generate their secure key in reality, the finite-key size effect\cite{christandl2009postselection,sheridan2010finite,ma2012statistical,curty2014finite} must be taken into account. In this section, we first extend the asymptoic SKR formula of Ref.\cite{cui2018phase} to non-asymptotic one against collective attack. Then based on the postselection technique developed in Ref.\cite{christandl2009postselection}, a formula against coherent attack is present.

\subsection {Security definition and SKR against collective attack}
As discussed above, in the end of NPP-TFQKD, Alice and Bob obtain a pair of bit string $S$ and $S'$ respectively. Ideally, the bit strings are secure and applicable to any cryptosystem if two fundamental conditions are met, namely correctness and secrecy. The correctness is, in simple terms, $S = S'$, which is guarantted by the error verification. The secrecy requires Eve's system $E$ is decoupled from Alice's key $S$, which is illustrated by $\rho_{SE} = U_S\otimes\rho_E$, where $\rho_{SE} = \sum_s(|s\rangle\langle s|\otimes\rho_E^s)$ denotes the density matrix of Alice and Eve's quantum state, $U_S = \sum_s\frac{1}{|S|}|s\rangle\langle s|$ denotes the uniform mixture of all possible value of $S$, $\{|s\rangle \}$ denotes the orthonormal basis of Alice's key $S$ and $\rho^s_E$ is Eve's the density matrix of Eve's system conditioned that Alice's key $S$ is in the state $|s\rangle$. Clearly, Alice's key $S$ is completely unknown to Eve in this ideal case. 

However, in finite-key size regime, the ideal condition $\rho_{SE} = U_S\otimes\rho_E$ usually can't be perfectly met. In Ref.\cite{muller2009composability}, a composoble secruity criteria is proposed. This criteria introduces secure parameters to describe some small probabilities of the keys $S$ and $S'$ varing from the ideal case. The protocol is $\epsilon_{cor}$-correct if $P(S \neq S')\leq\epsilon_{cor}$, i.e. the probability of $S \neq S'$ is less than $\epsilon_{cor}$. Similarly, the protocol is $\epsilon_{sec}$-secret if $\frac{1}{2}	\left \| \rho_{SE} - U_S\otimes \rho_E	\right \|_1 \leq \epsilon_{sec} $, which means $\rho_{SE}$ is $\epsilon_{sec}$ close to the ideal situation $U_S\otimes\rho_E$, where the symbol $\left \| M \right \|_1$ denotes trace norm of matrix $M$. In general, if a protocol is $\epsilon$-secure,   $\epsilon_{cor}+\epsilon_{sec}\le\epsilon$ must hold. To meet this criteria, with the same manner of Ref.\cite{sheridan2010finite}, the SKR formula of NPP-TFQKD against collective is given by

\begin{equation}
\begin{aligned}
R_{col} = &\frac{n}{N}\big{[}1  - \overline{I}_{AE}\big{]} - 
\frac{1}{N}\lambda_{EC} - \frac{1}{N}\lambda_{EV}- \frac{2}{N}log_2\frac{1}{\epsilon_{PA}} - \frac{7}{N}\sqrt{ n\  log_2(2/\epsilon_s) }\ ,
\label{SKR_collective}
\end{aligned}
\end{equation}
where $n = P_c^2Q_cN$ is the size of sifted key bits, $\overline{I}_{AE}$ is the upperbound of Eve's information on the sifted key bit if she launches collective attack, $\lambda_{EV}=log_2\frac{2}{\epsilon_{cor}}$ implies that $P(S\neq S')\leqslant \epsilon_{cor}$, $\epsilon_{PA}$ accounts for the probability of failure of privacy amplification, and $\epsilon_{s}$ measures the accuracy of the estimating the smooth min-entropy\cite{sheridan2010finite}. As shown in Ref.\cite{cui2018phase}, the estimation of $\overline{I}_{AE}$ against collectice attack depends on some experimentally observed parameters including the gains $Q_c$ and $Q_{xy}$. When the number of trials is finite, the expectations of these gains may vary from the experimentally observed values due to statistical fluctuations. Thus, another secure parameter $\epsilon_{PE}$\cite{curty2014finite} characterizing the  probablilty that parameter estimation fails must be taken into account. For instance, consider a set of $i.i.d.$ random variables $X_1X_2...X_N(X_i\in\{0,1\})$, the observed frequency of bit $1$ is usually not equal to its expectation $E(X)$, provided $N$ is finite. To solve this problem, we apply large deviation theory, specifically, the Chernoff bound to estimate a confidence interval of $X$ according to the obeserved value. In NPP-TFQKD, we can apply Chernoff bound\cite{curty2014finite,wang2017measurement,tang2014measurement} to estimate $Q_c$ and $Q_{xy}$ through the observed gains $\hat{Q}_{c}$ and $\hat{Q}_{xy}$ with a failure probability $\epsilon_{PE}$ respectively. For instance, we have that the expectation value of the gain $Q_{xy}$ satisfied ${Q}^-_{xy}\le Q_{xy}\le {Q}^+_{xy}$ with probability $1-\epsilon_{PE}$, where

\begin{equation}
\begin{aligned}
&{Q}_{xy}\ge{Q}^-_{xy}=\hat{Q}_{xy}(1+\frac{f(\epsilon_{PE}^{4}/16)} {\sqrt{N_{xy}\hat{Q}_{xy}}}), \\
&{Q}_{xy}\le{Q}^+_{xy}=\hat{Q}_{xy}(1-\frac{f(\epsilon_{PE}^{3/2}) }{\sqrt{N_{xy}\hat{Q}_{xy}}}),
\label{Chernoff_bound}
\end{aligned}
\end{equation}
, $f(\epsilon) = \sqrt{2ln(\epsilon^{-1})}$, and $N_{xy}$ denotes the total number of trails which Alice and Bob select decoy mode with intensity $x$ and $y$, respectively. As there are totally 11 gains to estimate in NPP-TFQKD \cite{cui2018phase}, the probability of occuring any failure in the estimations of 11 gains is $11\epsilon_{PE}$. Then applying the worst-case fluctuation analysis in the calculation of linear programming\cite{wang2017measurement,yu2015statistical}, we can bound $Y_{nm}(n+m\le 2)$, where $Y_{nm}$ is the probability of Charlie announcing a click message conditioned that Alice and Bob send Fock states $|n\rangle$ and $|m\rangle$ respectively. Furthermore, with Eq.(2) in Ref.\cite{cui2018phase}, we obtain $\overline{I}_{AE}$ with the failure probability of $11\epsilon_{PE}$.

Finally, Alice and Bob generate $NR_{col}$ bits secret key against collective attack with $\epsilon_{col}$-security. Obviousely, $\epsilon_{col}$ is not exceeding the sum of failure probabilities of error verification, privacy amplification, accuracy of smooth min-entropy and parameters estimation, say,
\begin{equation}
\epsilon_{col} \leq \epsilon_{PA}  + \epsilon_{cor} + \epsilon_{s} + 11\epsilon_{PE}.
\label{secure_para_col}
\end{equation}
Now we have introduced how to generate $\epsilon_{col}$-security keys against collective attack in NPP-TFQKD with finite-key effect. Next, we discuss how to obtain $\epsilon_{coh}$-security keys against coherent attack.

\subsection{Countermeasure of coherent attack}
According to Ref.\cite{christandl2009postselection}, it is proved that for a QKD protocol, the security against collective attack could be extended to be against coherent attack easily. We introduce the following corollary from the theorem 1 of Ref.\cite{christandl2009postselection} to tackle coherent attack in finite-key region.

 $Corollary$. The key rate $R_{coh}$ against coherent attack could be given by  
 \begin{equation}
 \begin{aligned}
 R_{coh} = R_{col}-\frac{126log_2(N+1)}{N},
 \label{SKR_coherent}
 \end{aligned}
 \end{equation}
while the key is $\epsilon_{coh}$-secure and

\begin{equation}
\begin{aligned}
\epsilon_{coh} = \epsilon_{col}(N+1)^{63}.
\label{secure_para_fun}
\end{aligned}
\end{equation}

 $Proof$. The proof is based on the theorem 1 of Ref.\cite{christandl2009postselection} and very similar proofs can be found in Ref.\cite{christandl2009postselection} and the appendix B of Ref.\cite{sheridan2010finite}. We denote $\mathcal{H_A}$, $\mathcal{H_B}$, and $\mathcal{M}$ are the Hilbert space of  Alice's ancilla $A$, Bob's ancilla $B$ and Clarlie's message $M$ respectively. Without compriomising the security, Charlie's messgage $M$ (click or not) can be treated as a quantum stated shared by Alice and Bob. The NPP-TFQKD protocol using Eq.\ref{SKR_collective} to generate keys could be viewed as a map $\mathcal{E}$ tranforming $A$, $B$ and $M$ into keys $S$ and $S'$ ($\big|S\big| = \big|S'
 \big|=NR_{col}$) respectively. Let $\mathcal{S}$ be a hypothecal map tranforming imperfect keys $S$ and $S'$  into perfect ones and define $\mathcal{F}=\mathcal{S}\circ\mathcal{E}$. Recall last subsection, it 
 asserts that $\lVert ((\mathcal{E}-\mathcal{F})\otimes id)\tau_{\mathcal{H}^N\mathcal{K}^N}\rVert_1\leqslant \epsilon_{col}$ holds when Eq.\ref{SKR_collective} is used to generate keys, where the de Finetti-Hilbert-Schmidt state $\tau_{\mathcal{H}^N\mathcal{K}^N}=\int\sigma^{\otimes N}_{\mathcal{H}\mathcal{K}}\mu(\sigma_{\mathcal{H}\mathcal{K}})$, $\mathcal{H}=\mathcal{H_A\otimes H_B\otimes M}$, $\sigma_{\mathcal{H}\mathcal{K}}$ is the pure state shared by Alice, Bob and Eve induced by any collective attack, and $\mu(\sigma_{\mathcal{H}\mathcal{K}})$ is the Haar measure on the pure state $\sigma_{\mathcal{H}\mathcal{K}}$.
   
 Next, we consider Eve may control another ancilla $R$ to obtain the purification $\tau_{\mathcal{H}^N\mathcal{K}^N\mathcal{R}}$ of $\tau_{\mathcal{H}^N\mathcal{K}^N}$. For such a purification,  $dim(\mathcal{R})$ is not larger than $(N+1)^{d^2-1}$\cite{christandl2009postselection} where $d=dim(\mathcal{H}\otimes\mathcal{K})=8$. Through controlling ancilla $R$, Eve's min-entropy on sifted key is decreased at most $2(d^2-1)log_2(N+1)$ bits. To meet the security, Alice and Bob may perform protocol $\mathcal{E}'$, in which privacy amplification shortens the sifted key into $NR_{col}-2(d^2-1)log_2(N+1)$ bits. Then we have $\lVert ((\mathcal{E}'-\mathcal{F}')\otimes id_{\mathcal{K}^N\mathcal{R}})\tau_{\mathcal{H}^N\mathcal{K}^N\mathcal{R}}\rVert_1\leqslant \epsilon_{col}$ still holds, where $\mathcal{F}'$ is a hypothecal map generating perfect keys. 
 
 Finally, we apply the theorem 1 of Ref.\cite{christandl2009postselection} and obtain
 
 \begin{equation}
 \begin{aligned}
 \lVert ((\mathcal{E}'-\mathcal{F}')\otimes id_{\mathcal{K}^N\mathcal{R}})\rho_{\mathcal{H}^N\mathcal{K}^N\mathcal{R}}\rVert_1\leqslant\lVert ((\mathcal{E}'-\mathcal{F}')\otimes id_{\mathcal{K}^N\mathcal{R}})\tau_{\mathcal{H}^N\mathcal{K}^N\mathcal{R}}\rVert_1\leqslant \epsilon_{col}(N+1)^{d^2-1}.
 \nonumber
 \end{aligned}
 \end{equation}
Since $\rho_{\mathcal{H}^N\mathcal{K}^N\mathcal{R}}$ is any state shared by Alice, Bob and Eve, this inequality clearly shows that the protocol $\mathcal{E}'$ is $\epsilon_{col}(N+1)^{d^2-1}$-secure for any coherent attack. Substituting $d=8$, we end the proof.\hfill $\square$

According to the corollary, if Alice and Bob want to generate $\epsilon_{coh}$-secure keys against any attack, they will calculate the parameter $\epsilon_{col}$ with Eq.\ref{secure_para_fun}, and generate keys with the fromulae Eqs.\ref{SKR_coherent}, \ref{SKR_collective} and \ref{secure_para_col}. 

To evaluate the performance of NPP-TFQKD in finite-key region, simulations in fiber channel are performed here. We assume the dark-count rate of SPD is $10^{-10}$ per trial, the detection efficiency is $14.5\%$ and optical misalignment is $1.5\%$. The attenuation of fiber is $0.2dB/km$ and the fiber tranmittance is $10^{-0.2L/10}$ where $L$ is fiber length. The total secure parameter $\epsilon_{coh}$ in Eq.\ref{secure_para_fun} is fixed as $10^{-10}$. 
In addition to fixed parameters above, there are some parameters should be optimized to maximize the SKR. There are 10 parameters should be optimized in total. The first set is decoy intensities $\mu$, $\nu$ and $\omega$. The second set is probabilities of modes and intensities. $P_c$ denote probabilities of choosing code mode and $P_d^\mu$ $P_d^\nu$ $P_d^\omega$ denote probabilities of choosing decoy mode with intensity $\mu$, $\nu$, $\omega$. It worth noting that probabilities of vacuum state is $P_d^o =1 - P_c - P_d^\mu - P_d^\nu - P_d^\omega$. The number of pulses they both select code mode is $NP_c^2$ and they select decoy mode with intensity $x$ and $y$ respectively is $NP_d^x P_d^y$. The other set is $\epsilon_{PA}$, $\epsilon_{cor}$, $\epsilon_{s}$ and $\epsilon_{PE}$ satisfying Eq.\ref{secure_para_col}. Define $\epsilon_{cor} = \epsilon_{col} r_{cor}$, $\epsilon_{cor} = \epsilon_{col} r_{cor}$,  $\epsilon_{s} = \epsilon_{col} r_{s}$ and $\epsilon_{PE} = \epsilon_{col} (1 - r_{sec} - r_{cor}- r_s)/11$.

The optimized paramters can be regarded as a vector $\vec{v} = [\mu, \nu, \omega,P_c, P_d^\mu, P_d^\nu,P_d^\omega,r_{sec},r_{cor},r_{s}]$. Noting that the convex form of function $R_{coh} = F(\vec{v})$\cite{Lu2019parameter,wang2018machine} is not guaranteed, we choose particle swarm optimization algorithm(PSO) which can optimize the non-smooth function and non-convex function\cite{kennedy2010particle} to search the best $\vec{v}$ to maximize the $R_{coh}$.

\begin{figure}[htbp]
\includegraphics[width=9cm]{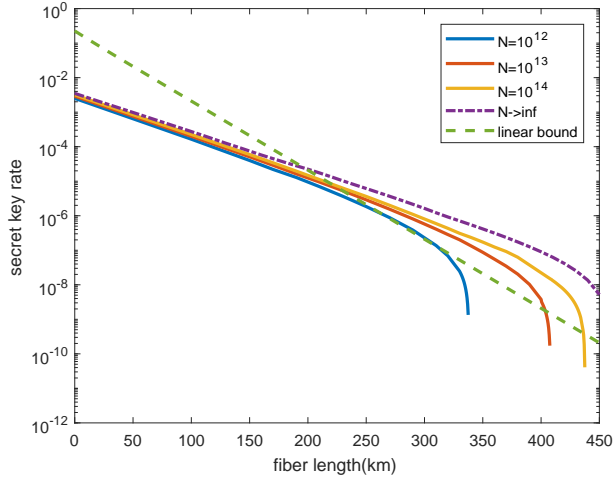}
\caption{\label{finitKey_SKR_L} SKR versus distance between Alice and Bob for three different pulses number ($N=12$: blue, $N=13$: red, $N=14$: yellow). The purple dot-dash line is asymptotic SKR and the green dash line is the linear bound}
\end{figure}

\begin{figure}[htbp]
\includegraphics[width=9cm]{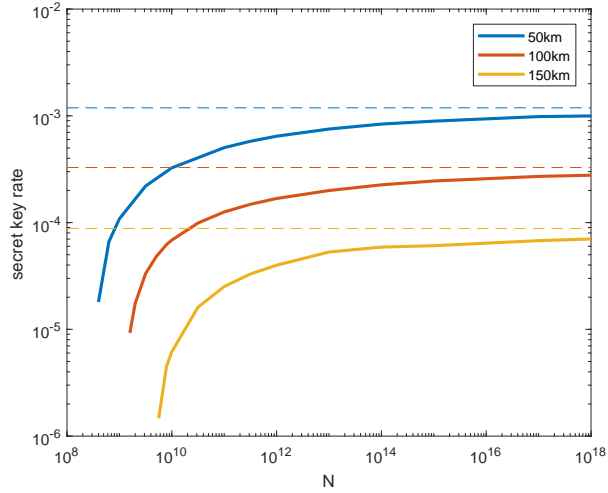}
\caption{\label{finitKey_SKR_N} Secret key rate in logarithmic scale as a function of pulses number $N$ for three different distance between Alice and Bob($50km$: blue, $100km$: red, $150km$: yellow). The solid lines denote non-asymptotic SKR and the dash lines show corresponding asymptotic SKR}
\end{figure}

 The results of the simulations are illustrated in Fig.\ref{finitKey_SKR_L} and \ref{finitKey_SKR_N}. In Fig.\ref{finitKey_SKR_L}, we fix the pulses number $N$ to be $10^{12}$, $10^{13}$, $10^{14}$ and simulate the SKR as a function of distance between Alice and Bon. In Fig.\ref{finitKey_SKR_N} the distance is fixed to be $50km$, $100km$ and $150km$, then we simulate the SKR as a function of $N$. The results show that compared with asymptotic situation, the protocol still works well in non-asymptotic situations and the linear bound is still overcomed when $N \ge 10^{12}$.

\section{NPP-TFQKD with both large random intensity fluctuation and finite-key size effect}

Except for finite-key size effect, a ubiquitous loophole in practical QKD system is intensity fluctuation\cite{zhou2017implementing,wang2009decoy,wang2007decoy}. When applying decoy state technique, accurate intensity values are required to ensure the correct estimation of $Y_{nm}$\cite{cui2018phase}. However, it's very difficult to control the intensity of WCP exactly in practical QKD system since noise, time jitter, problem of modulation and other imperfections of devices. It brings potential loopholes and may allow Eve to perform sophisticated attacks. In this section, we discuss the NPP-TFQKD with large random intensity fluctuation in finite-key size regime. The main contribution of this section is that we present a countermeasure of both large random intensity fluctuation and finite-key size effect of NPP-TFQKD. By applying our method, the NPP-TFQKD with large random intensity fluctuation can remain its advantage of breaking the linear bound.

\subsection{Analytical formula of 4-intensity decoy state method of NPP-TFQKD}

Before proposing the intensity fluctuation model of NPP-TFQKD, we will introduce our analytical formula of 4-intensity decoy state method. In 'Parameter estimation and privacy amplification' step, the n-photon yield can be estimated by linear programming or analytical formula\cite{wang2013three,yu2015statistical}. However, the analytical formula of NPP-TFQKD is not given. In our countermeasure of imperfect WCP source loophole in next section, the analytical formula is needed. To make the NPP-TFQKD more practical, the analytical formula of 4-intensity decoy state method is proposed.

Define $q_{nm} = p_n^\mu p_m^\mu Y_{nm}$ where $p_n^{x} = e^{-x}\frac{x^{n}}{n!}$. To estimate the upper bound of $I_{AE}$, we have to estimate the upper bound of $q_{00}$, $q_{10}$, $q_{01}$, $q_{20}$, $q_{02}$, $q_{11}$ and lower bound of $q_{sum} = q_{00}+q_{10}+q_{01}+q_{20}+q_{02}+q_{11}$. The upper and lower bound of $Y_{nm}$ can be estimated by applying linear programming whose constraints are joint of:
\begin{equation}
Q_{xy} = \sum_{n=0}^\infty\sum_{m=0}^\infty{p_n^xp_m^yY_{nm}}\ \ (x,y \in \{\mu,\nu,\omega,o\}),
\label{Eq_QY}
\end{equation}
where $ 0\le Y_{nm}\le 1$. Noting that these $p_n^{x}$ depend on the intensity $x$, it's obvious that the $p_n^{x}$ in Eq.(\ref{Eq_QY}) are uncertain and the linear programming will be not valid any more if we can't control intensities exactly. Intuitively, we can still get secure bound of key rate if we correctly replace coefficients $p_n^{x}$ by its upper and lower bound in analytical formula. Thus we present an analytical formula before building the fluctuation model.

We will use superscript or subscript $'+'$ and $'-'$ to express, respectively, upper and lower bound of a variable and we denote intensities in decoy mode by $\mu$ ,$\nu$, $\omega$, $o$ where $\mu > \nu > \omega >o$ and $o$ is the vacuum state. It is worth noting that, the intensity of code mode should be the same as one of $\mu$ ,$\nu$ or $\omega$. To make our formula more clear, we suppose code mode intensity is $\mu$ and denote $p_n^\mu$ $p_n^\nu$ $p_n^\omega$ by $a_n$, $b_n$, $c_n$.

Here we will demonstrate how to estimate n-photon yield by analytical formulas as follows. The details are showed in Appendix.

\textbf{Estimation of $q_{00}$:}

$q_{00} = a_0^2Q_{oo}$.
\\

\textbf{Upper bound of $q_{10}$ and $q_{01}$:} 

\begin{equation}
{q}^+_{10} = a_0a_1\frac{K_2(c_2a_3 - c_3a_2)+K_1(b_2c_3 - b_3c_2)+K_3(a_2b_3-a_3b_2)}{b_2(a_1c_3 - c_1a_3)+b_1(c_2a_3 - c_3a_2)+b_3(a_2c_1 - a_1c_2)},
\end{equation}
where $K_1 = Q_{\mu o} - a_0Q_{oo}$, $K_2 = Q_{\nu o} - b_0Q_{oo}$ and $K_3 = Q_{\omega o} - c_0Q_{oo}$.

Expression of ${q}^+_{01}$ is similar, the difference is $K_1 = Q_{o \mu} - a_0Q_{oo}$, $K_2 = Q_{o\nu} - b_0Q_{oo}$, $K_3 = Q_{o\omega} - c_0Q_{oo}$.
\\

\textbf{Upper bound of $q_{20}$ and $q_{02}$:} 

\begin{equation}
{q}^+_{20} = a_0a_2\frac{H_1c_1 - L_3a_1}{a_2c_1 - a_1c_2},
\end{equation}
where $H_1 = Q_{\mu o} - a_0Q_{oo}$, $L_3 = Q_{\omega o} - c_0Q_{oo} - (\sum_{n=3}^\infty{c_n})$. Formula of ${q}^+_{02}$ is similar but  $H_1 = Q_{o\mu} - a_0Q_{oo}$, $L_3 = Q_{o\omega} - c_0Q_{oo} - (\sum_{n=3}^\infty{c_n})$.
\\

\textbf{Upper bound of $q_{11}$:} 

\begin{equation}
{q}^+_{11} = {Q_{\mu\mu} - a_0(Q_{\mu o}+Q_{o\mu}) +a_0^2Q_{oo}}.
\end{equation}

\textbf{Lower bound of $q_{sum}$:}

We take two steps to calculate the ${q}^+_{sum}$. Define $q_{sum} = q_{t_1} + q_{t_2}$, where $q_{t_1} = q_{00} + q_{10} + q_{01} + q_{20} + q_{02} + q_{11}$ and $q_{t_2} = q_{11}$.

\begin{equation}
\begin{aligned}
&q_{t_1}^- = a_0[Q_{o\mu} + Q_{\mu o} - 2\sum_{n=3}^\infty{a_n}] - a_0^2Q_{oo}\ ,\\
&q_{t_2}^- = (a_1)^2\frac{T_1b_1b_2 - T_2a_1a_2}{a_1^2b_1b_2 - b_1^2a_1a_2}\ ,
\end{aligned}
\end{equation}
where $T_1 = Q_{\mu\mu} - a_0(Q_{o\mu}+Q_{\mu o})+a_0^2Q_{oo}$ and $T_2 = Q_{\nu\nu} - b_0(Q_{o\nu}+Q_{\nu o})+b_0^2Q_{oo}.$

The lower bound of $q_{sum}$ is
\begin{equation}
q_{sum}^- = q_{t_1}^- + q_{t_2}^-.
\end{equation}

\subsection {Estimation of average intensity}
In this subsection, we briefly introduce the simple tomography technique proposed by Ref.\cite{wang2007decoy}. Based on this work, we propose a large random intensity fluctuation model in finite-key size regime.

\begin{figure}[htbp]
\includegraphics[width=9cm]{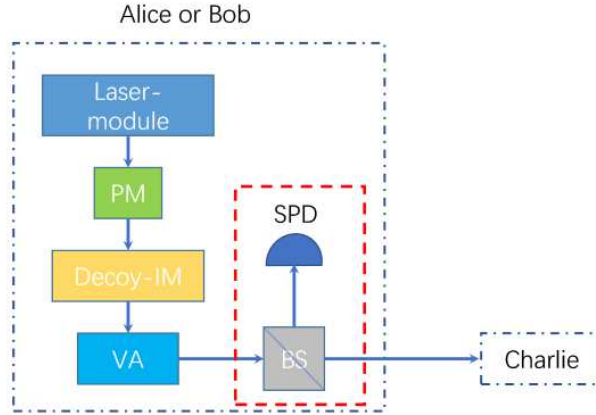}
\caption{\label{tomography} The main structure of simple tomography technique is showed in the red dash line box. Alice (Bob) does all operations as usual except attenuate intensity to $2x$ when she (he) actually wants $x$, then she (he) splits it by a 50:50 BS. One of the pulse is sent to Charlie and the other is measured by a local low dark-count SPD. By observing the count rate of the local SPD, Alice (Bob) can estimate the bound of average intensity. PM denotes phase modulator, decoy-IM denotes intensity modulator. VA denotes variable attenuator. BS denotes 50:50 beam splitter and SPD denotes single photon detector.} 
\end{figure}

As illustrated in Fig.\ref{tomography}, Alice (Bob) should firstly produce a WCP with intensity $2x$ when she(he) actually wants $x$. Before sending the WCP to Charlie, she (he) splits it by a 50:50 BS. One of the pulse is sent to Charlie and the other one is measured by a local low dark-count SPD whose detection efficiency is $\eta$.
After sending $N_x$ $x$-intensity WCPs, the local detector's count number is $n_x$ where the dark count is ignored since it's orders of magnitude lower than light count.
Because of the random fluctuations, whenever Alice (Bob) wants to modulate intensity $x$, she (he) actually modulates $x_i = (1 +\delta_i)\bar{x}$, where $\bar{x}$ is the average intensity and the instantaneous fluctuation $\delta_i$ is an unknow value. Mathematically, the click rate $h_x$ is:
\begin{equation}
 h_x = \sum^{N_x}_{i=0}(1 - e^{-\eta x_i})/N_x.
\label{ave_inten}
\end{equation}

As proof in Ref.\cite{wang2007decoy}, the upper and lower bound of $\bar{x}$ is:
\begin{equation}
\begin{aligned}
&\bar{x}\le x_+ = \frac{ 1 - \sqrt{ 1 - 2h_x(1 + \zeta)} }{\eta(1 + \zeta)}\ ,  \\
&\bar{x}\ge x_- = h_x/\eta + h_x^2/2\eta - \eta^2x_+^3/3!\ ,
\label{bound_of_ave_inten}
\end{aligned}
\end{equation}
where $\zeta = \sum\delta_i^2/N_x\le( Max\{|\delta_i|\})^2$.

However, this conclusion in Ref.\cite{wang2007decoy} can't be used in non-asymptotic situations directly. Here we apply large deviation theory to make the method met practice. 

Noting that the distribution of intensity fluctuation is not independent identically distributed in most cases, we choose Azuma's inequality\cite{yin2010security,boileau2005unconditional,azuma1967weighted} rather than Chernoff bound to estimate the confidence interval of $h_x$. When the observed count number is $\hat{n}$, The upper and lower bound of $h_x$ is:
\begin{equation}
\begin{aligned}
&h_x\le h_x^+ = (\hat{n} + \sqrt{2\hat{n} ln\frac{1}{\epsilon_h}})/N_x\ ,\\
&h_x\ge h_x^- = (\hat{n} - \sqrt{2\hat{n} ln\frac{1}{\epsilon_h}})/N_x\ ,
\label{bound_of_h}
\end{aligned}
\end{equation}
where the $\epsilon_h$ is secure parameter of estimation. Then the bound of the average intensity is corrected as
\begin{equation}
\begin{aligned}
&\bar{x}\le x_+ = \frac{ 1 - \sqrt{ 1 - 2h_x^+(1 + \zeta)} }{\eta(1 + \zeta)}\ ,  \\
&\bar{x}\ge x_- = h_x^-/\eta + (h_x^-)^2/2\eta - \eta^2x_+^3/3!\ .
\label{corrected_ave_inten}
\end{aligned}
\end{equation}

\subsection {Model of NPP-TFQKD with both intensity fluctuation and finite-key size effect}

In this subsection, we will propose our countermeasure model.
Firstly, we should define some symbols. Let's take the first decoy state(intensity is $\mu$) as an example.
When we want sent $\mu$-intensity weak coherent pulse, we actually prepare $\mu_i = \bar{\mu}(1 + \delta_i)$ since the intensity fluctuation. The intensity range is $\mu^{\pm} = \bar{\mu}(1 + \delta^{\pm})$ where $\delta^{+} = Max\{\delta_i\}$ and $\delta^{-} = Min\{\delta_i\}$.

With definitions above, the density matrix of the source with fluctuation can be describe by:
\begin{equation}
\rho'_{\mu} = \sum_{i=1}^{N_\mu}\sum_{n=0}^{\infty}{e^{-\mu_i}\frac{\mu_i^n}{n!}|n\rangle\langle{n}|}/N_{\mu},
\label{fluc_density_mat}
\end{equation}
and the $a_n$ is re-written as:
\begin{equation}
\begin{aligned}
a_n = \frac{1}{{N_\mu}}\sum_{i=1}^{N_\mu}e^{-\mu_i}\frac{\mu_i^n}{n!}=
a_n = \frac{\bar{\mu}^ne^{-\bar{\mu}}}{n!{N_\mu}}\sum_{i=1}^{N_\mu}e^{-\delta_i\bar{\mu}}( 1+\delta_i)^n.
\label{fun_of_an}
\end{aligned}
\end{equation}

By applying taylor expansion to Eq.(\ref{fun_of_an}), we get:
\begin{equation}
\begin{aligned}
a_n &= \frac{\bar{\mu}^ne^{-\bar{\mu}}}{n!{N_\mu}}\sum_{i=1}^{N_\mu}(1 - \delta_i\bar{\mu} + \frac{(\delta_i\bar{\mu})^2}{2!} -...)( 1 + C_n^1\delta_i + C_n^2(\delta_i)^2+...)
\\&=\frac{\bar{\mu}^ne^{-\bar{\mu}}}{n!{N_\mu}}[1+\sum_{i=1}^{N_\mu}(n-\bar{\mu})\delta_i + o(\delta_i)].
\label{boundEq1}
\end{aligned}
\end{equation}
Noting an important fact that $\sum_{i=1}^{N_\mu}\delta_i = 0$, we find the first order item of $\delta$ is not exist in Eq.(\ref{boundEq1}). i.e, the Eq.(\ref{boundEq1}) can be re-written as: 
\begin{equation}
\begin{aligned}
a_n = \frac{\bar{\mu}^ne^{-\bar{\mu}}}{n!{N_\mu}} [\sum_{i=1}^{N_\mu}e^{-\delta_i\bar{\mu}}( 1+\delta_i)^n - \sum_{i=1}^{N_\mu}(n-\bar{\mu})\delta_i].
\label{boundEq2}
\end{aligned}
\end{equation}

Define function $f_n(\delta) = e^{-\delta\mu^+}(1+\delta)^n - (n-\mu^+)\delta$, $\delta\in[\delta_- ,\delta_+]$. $f_n(\delta^u)$ and $f_n(\delta^l)$ are, respectively, maximum and minimum values which can be easily found by optimization algorithms in interval $[\delta_- ,\delta_+]$. $\delta^u$ and $\delta^l$ are, respectively, maximum and minimum value points. 

Noting that the function $g(\mu) = \frac{{\mu}^ne^{-{\mu}}}{n!}$ is monotonically increasing function when $n\ge 1$ and $\delta\in[-1,1]$, when $n\ge0$, we can obtain a tight bound of $a_n$ as 
\begin{equation}
\begin{aligned}
&{a}_n \le {a}_n^+ = \frac{(\mu^+)^ne^{-\mu^+}}{n!}f(\delta^u),\\
&{a}_n \ge {a}_n^- = \frac{(\mu^-)^ne^{-\mu^-}}{n!}f(\delta^l).
\label{bound_of_an}
\end{aligned}
\end{equation}
Espacially, when $n = 0$,
\begin{equation}
\begin{aligned}
&a_0 \ge a_0^- = e^{-\mu^+}; \ \ \ a_0 \le a_0^+ = e^{-\mu^-}.
\label{bound_of_a0}
\end{aligned}
\end{equation}

However, without introduction of average intensity, when $n \ge 1$:
\begin{equation}
\begin{aligned}
&a_n \ge a_n^- = \frac{e^{-\mu^-}(\mu^-)^n}{n!}, \ \ \ a_n \ge a_n^+ = \frac{e^{-\mu^+}(\mu^+)^n}{n!}.
\end{aligned}
\end{equation}
and when $n = 0$:
\begin{equation}
a_0 \ge a_0^- = e^{-\mu^+}, \ \ \ a_0 \le a_0^+ = e^{-\mu^-}. \\
\end{equation}

\begin{figure}[htbp]
\includegraphics[width=8cm]{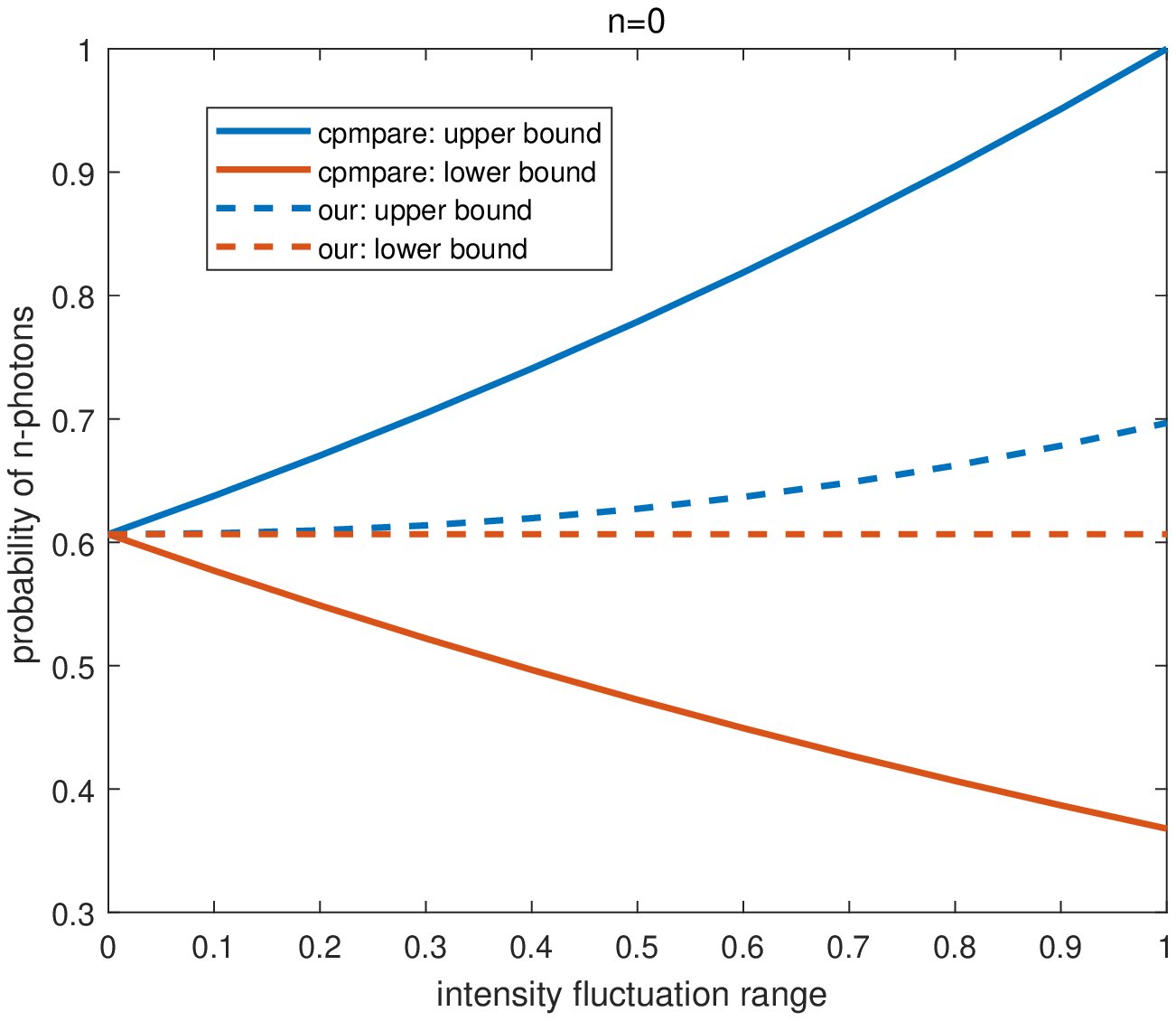}
\includegraphics[width=8cm]{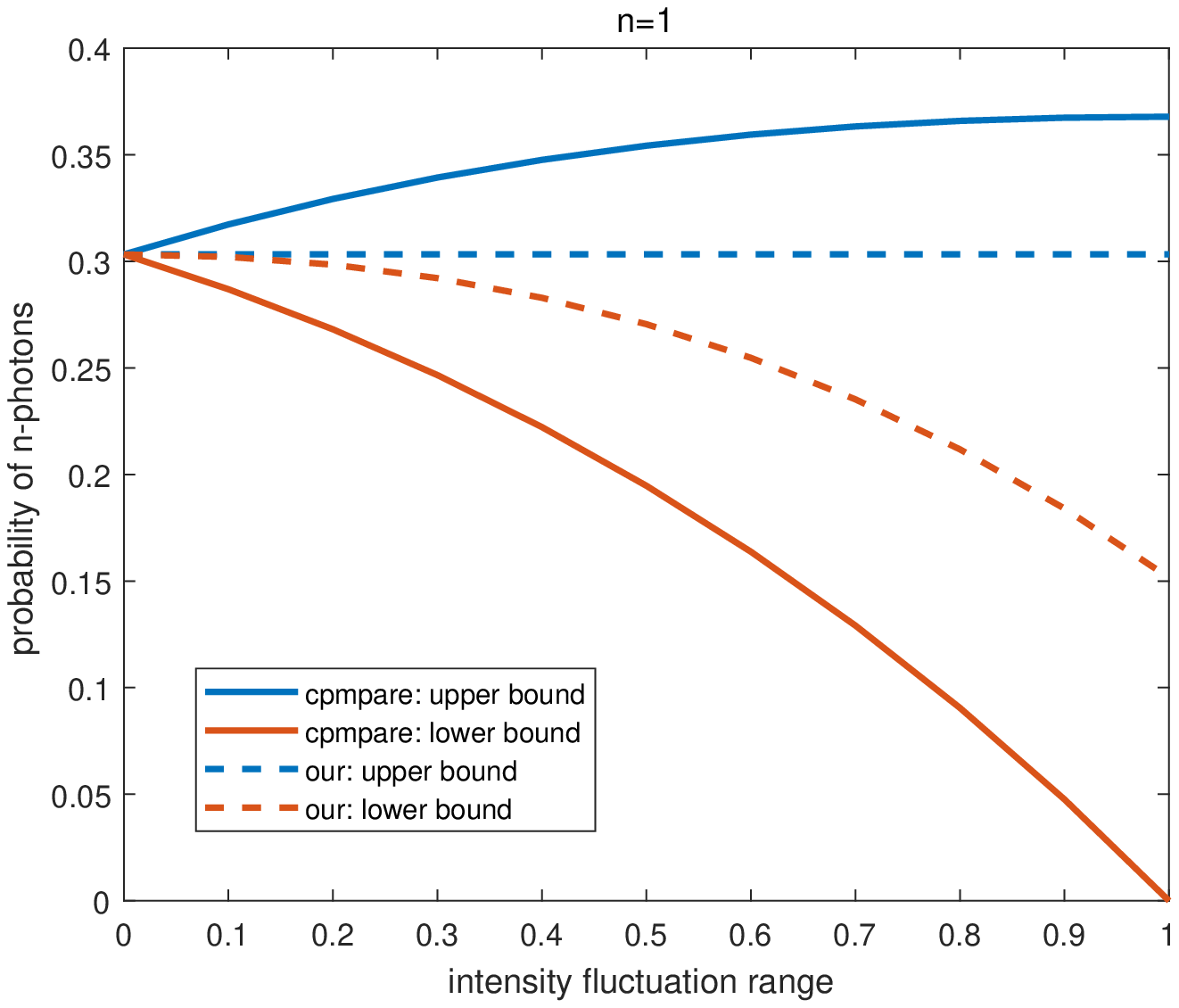}
\includegraphics[width=8cm]{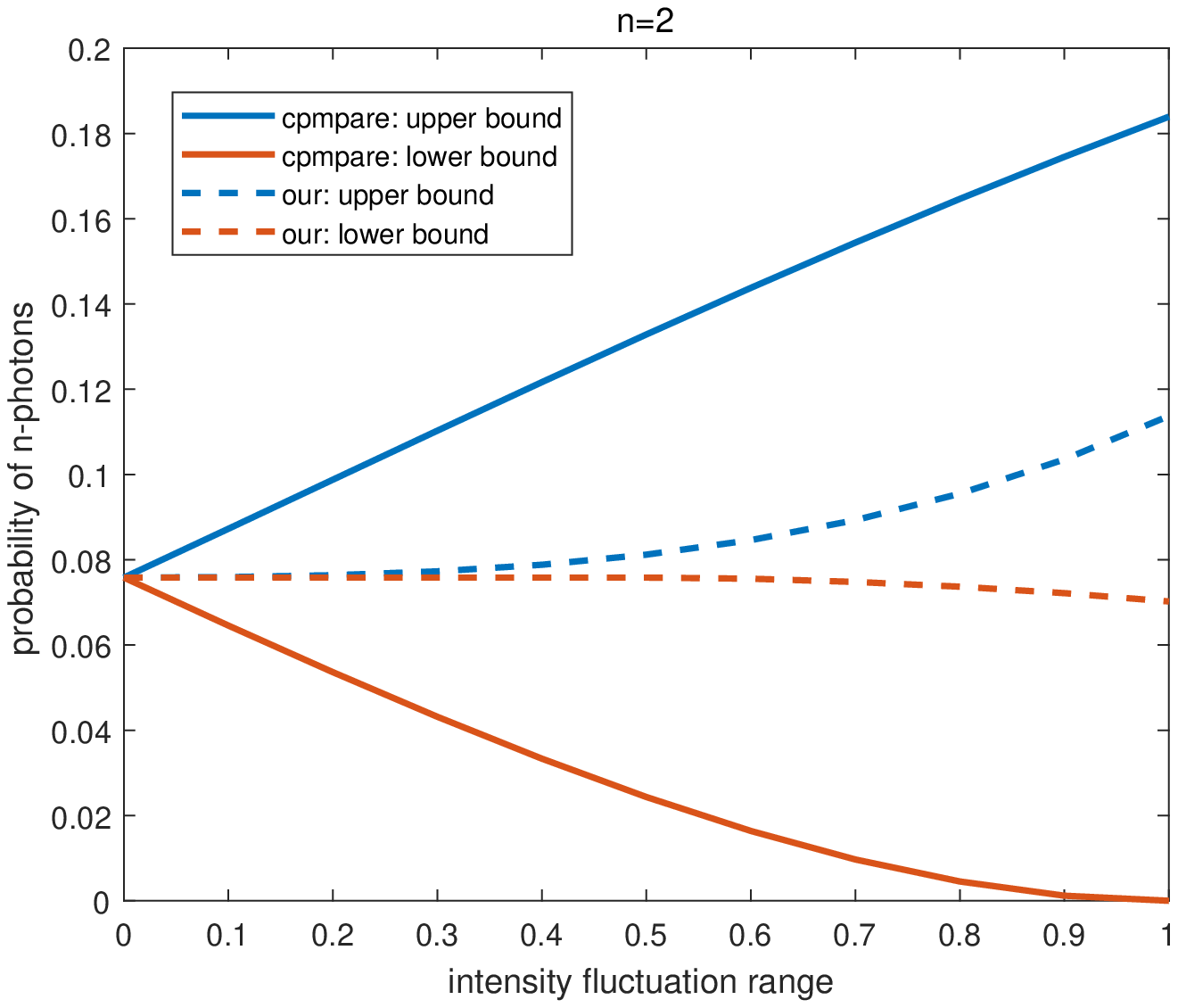}
\includegraphics[width=8cm]{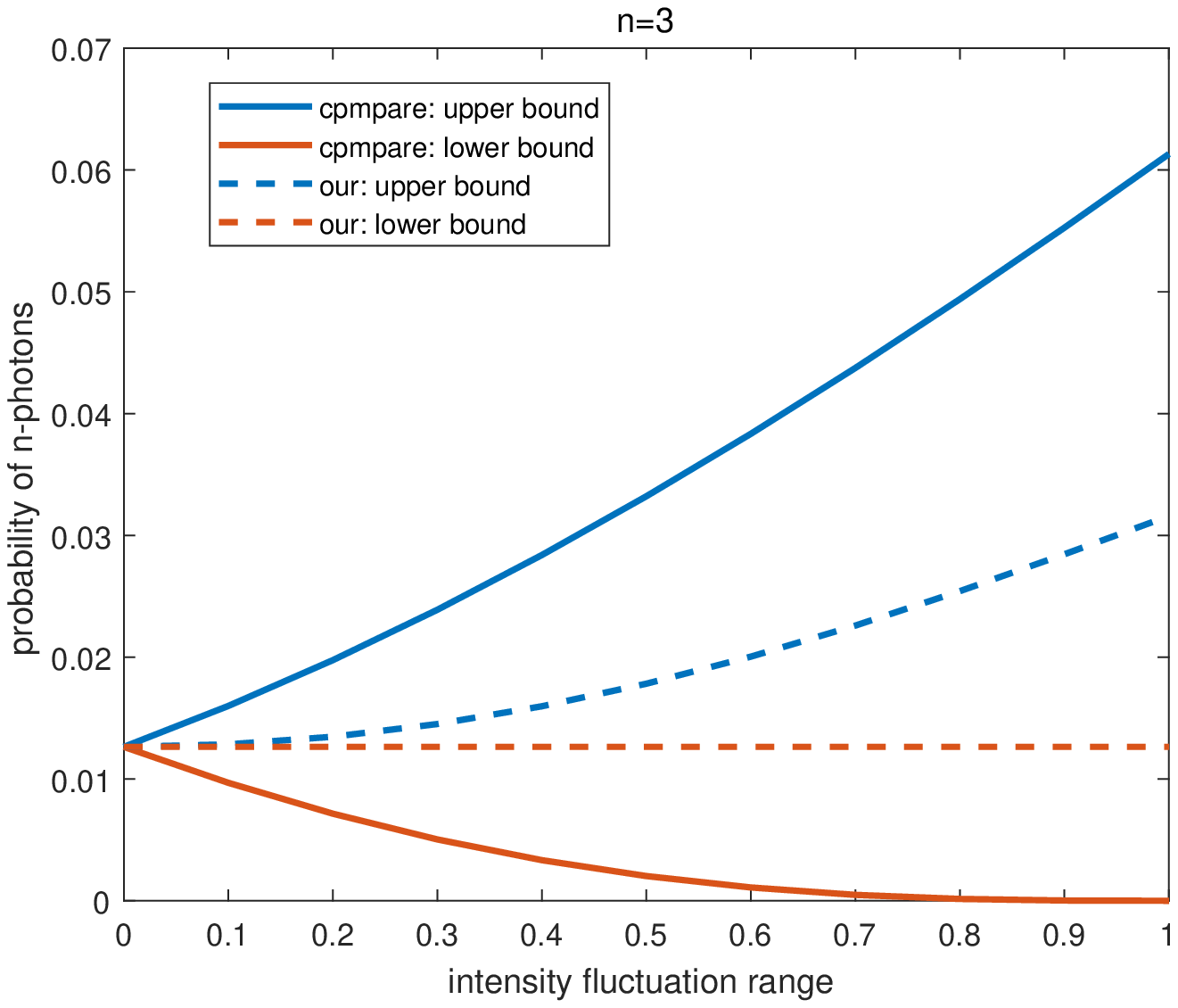}
\caption{
\label{pn}  We compare the upper and lower bound of $a_n$ with two methods. The intensity is fixed at $0.5$ and $n=0,1,2,3$ in different figures. In these figures, the blue and red lines denote the upper bound and lower bound respectively. The solid lines denote the model only considering fluctuation range and dash lines are our method which introducing the average intensity. It's obvious that our method estimates the bound much tighter.
}
\end{figure}

The difference between introducing average intensity or not is showed in Fig.\ref{pn},
it indicates that the introduction of average intensity can significantly tighten the bound.

By substituting Eq.(\ref{Chernoff_bound}),(\ref{bound_of_an}) and (\ref{bound_of_a0}) into our analytical formula, we can obtain bounds of $q_{nm}$ in uncertain intensity and finite-key size regime.

\textbf{Upper bound of $q_{00}$:}

 $q_{00} = ({a}^+_0)^2Q_{oo}$;
\\

\textbf{Upper bound of $q_{10}$ and $q_{01}$:} 

We take ${q}^+_{10}$ as example, the ${q}^+_{01}$ is similar.

\begin{equation}
\begin{aligned}
{q}^+_{10} = {a}^+_0{a}^+_1\frac{{K}^+_2({c}^+_2{a}^+_3 - {c}^-_3{a}^-_2)-{K}^-_1({b}^+_3{c}^+_2 - {b}^-_2{c}^-_3)-{K}^-_3({a}^+_3{b}^+_2-{a}^-_2{b}^-_3)}{{b}^-_2({c}^-_1{a}^-_3 - {a}^+_1{c}^+_3)-{b}^+_1({c}^-_2{a}^-_3 - {c}^+_3{a}^+_2)+{b}^-_3({a}^-_2{c}^-_1 - {a}^+_1{c}^+_2)},
\end{aligned}
\end{equation}
where ${K}^-_1 = Q_{\mu o}^- - {a}^+_0Q_{oo}^+$, ${K}^+_2 = Q_{\nu o}^+ - {b}^-_0Q_{oo}^-$ and ${K}^-_3 = Q_{\omega o}^- - {c}^+_0Q_{oo}^+$.
\\

\textbf{Upper bound of $q_{20}$ and $q_{02}$:} 

We take ${q}^+_{20}$ as example.
\begin{equation}
\begin{aligned}
{q}^+_{20} = {a}^+_0{a}^+_2\frac{{H}^+_1{c}^+_1 - {L}^-_3{a}^-_1}{{a}^-_2{c}^-_1 - {a}^+_1{c}^+_2},
\end{aligned}
\end{equation}
where ${H}^+_1 = Q_{\mu o}^+ - {a}^-_0Q^-_{oo}$, ${L}^-_3 = Q_{\omega o}^- - {c}^+_0Q_{oo}^+ - (\sum_{n=3}^\infty{{c}^+_n})$.
\\

\textbf{Upper bound of $q_{11}$:} 

\begin{equation}
\begin{aligned}
{q}^+_{11} = {Q_{\mu\mu}^+ - {a}^-_0(Q_{\mu o}^-+Q_{o\mu}^-) + {a}^+_0 2Q_{oo}^+}.
\end{aligned}
\end{equation}

\textbf{Lower bound of $q_{sum}$:}

\begin{equation}
\begin{aligned}
&{q}^-_{sum} = {q}^-_{t1} + {q}^-_{t2}\ , \\
&{q}^-_{t1} = {a}^-_0[Q^-_{o\mu} + Q^-_{\mu o} - 2\sum_{n=3}^\infty{{a}^-_n}  - {a}^+_0Q_{oo}^+]\ ,\\
&{q}^-_{t2} = \frac{{T}^-_2{a}^-_1{a}^-_2 - {T}^+_1{b}^+_1{b}^+_2}{(a^+_1b^+_1)(a_2^+b_1^+ - a_1^-b_2^-)}\ ,
\end{aligned}
\end{equation}
  where ${T}^+_1 = Q^+_{\mu\mu} - {a}^-_0(Q^-_{o\mu}+Q^-_{\mu o}-{a}^+_0Q^+_{oo})$ and ${T}^-_2 = Q^-_{\nu\nu} - {b}^+_0(Q^+_{o\nu}+Q^+_{\nu o}-{b}^-_0Q^-_{oo}).$


\begin{figure}[htbp]
\includegraphics[width=9cm]{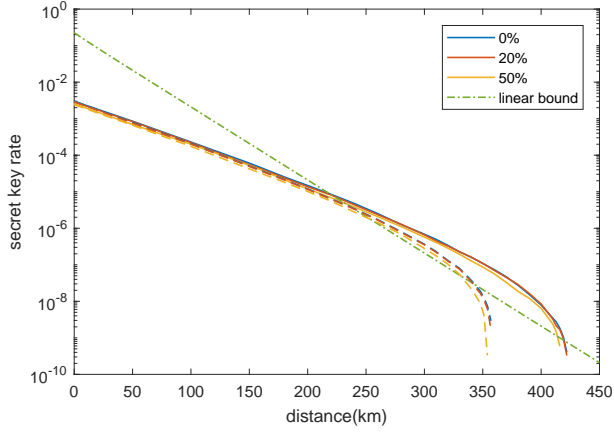}
\caption{\label{intenFluc_SKR_L} Asymptotic secret key rate in logarithmic scale as a function of distance between Alice and Bob for different fluctuations and pulses number($\delta^{\pm} = \pm 0\%$: blue, $\delta^{\pm} = \pm 20\%$: red, $\delta^{\pm} = \pm 50\%$: yellow. $N= 10^{14}$: solid lines, $N= 10^{13}$: dash lines). The green dot-dash line denotes linear bound.}
\end{figure}

The simulation of NPP-TFQKD with both large random intensity fluctuation and finite-key size effect is shown in Fig.\ref{intenFluc_SKR_L}. We fix pulse number, dark-count rate, detection efficiency, misalignment, total secure parameter $\epsilon_{coh}$ and secure parameter of Azuma's inequality in Eq.(\ref{bound_of_h}) $\epsilon_h$ to $10^{14}$, $10^{-10}$ ,$14.5\%$, $1.5\%$, $10^{-10}$ and $10^{-10}$ respectively. To emphasize the countermeasure of intensity fluctuation, we simulate the SKR as a function of distance for different intensity fluctuation range and optimize SKR by PSO algorithm as introduced in Sec.\textrm{III}. The simulation result in Fig.\ref{intenFluc_SKR_L} indicates that by applying our countermeasure model, the large random intensity fluctuation has very limited influence on the performance of NPP-TFQKD.


\section{Conclusion}
In this article, we have discussed some practical issues of NPP-TFQKD based on Ref.\cite{cui2018phase}. We firstly analyzed the issue of finite-key size effect and solve this problem by applying post-selection technique for quantum channels\cite{christandl2009postselection} and using Chernoff Bound to estimate statistic fluctuations of observed values. The simulation shows that NPP-TFQKD works well in non-asymptotic situations.

Another contribution of this work is we propose a countermeasure of intensity fluctuation. We introduce an analytical formula of decoy state method to meet the needs of our fluctuation model. Then we propose our intensity fluctuation model to deal with large random intensity fluctuation problem in the source side. Our model is practical since it doesn't need any extra information except average intensity and fluctuation range. Our simulation results suggest that by applying our method,the non-asymptotic SKR can still break the linear bound even if the large random intensity fluctuation is taken into account.

This work has been supported by the National Key Research and Development Program of China (Grant No. 2016YFA0302600), the National Natural Science Foundation of China (Grant Nos. 61822115, 61775207, 61702469, 61771439, 61622506, 61627820, 61575183), National Cryptography Development Fund (Grant No. MMJJ20170120) and Anhui Initiative in Quantum Information Technologies.


\section*{Appendix: proof of analytical formula}

Firstly, we should introduce an important conclusion\cite{wang2013three,yu2015statistical}. $|\mu\rangle$ and $|\nu\rangle$ are coherent states, intensity $\mu$ is larger than $\nu$. When $m>n$, there is:

\begin{equation}
\frac{p_m^\mu}{p_m^\nu}\geq\frac{p_n^\mu}{p_n^\nu}\ \ (\mu\geq\nu \ \ ;\ \  m\geq n). 
 \label{Beq0}
\end{equation}

\textbf{Upper bound of $q_{00}$:} 

The $Y_{00} = Q_{oo}$ since $o$ is vacuum state. Thus:
\begin{equation}
q_{00} = a_0^2Y_{00} = a_0^2Q_{oo}.
\end{equation}

\textbf{Upper bound of ${q}_{01}$ and ${q}_{10}$:} 

We take ${q}^+_{01}$ as an example and the proof of ${q}^+_{10}$ is similar. Define $K_1 = Q_{o\mu} - a_0Q_{oo}$, $K_2 = Q_{o\mu} - b_0Q_{oo}$, $K_3 = Q_{o\mu} - c_0Q_{oo}$. Noting that $Q_{ox} - p_0^xQ_{oo} = p_1^xY_{01}+p_2^xY_{02}+p_2^xY_{02}+....$, We get an equation set:

\begin{equation}
\left\{
\begin{aligned}
&K_1  =  a_1Y_{01} +  a_2Y_{02} +  a_3Y_{03} + ....\\ 
&K_2  =  b_1Y_{01} +  b_2Y_{02} +  b_3Y_{03} + ....\\ 
&K_3  =  c_1Y_{01} +  c_2Y_{02} +  c_3Y_{03} + ....\\ 
\end{aligned}
\right.
\label{Beq1}
\end{equation}.

By applying conclusion(\ref{Beq0}) and defining $\tau = \sum^\infty_{k=3}b_kY_{0k}$, we can derive inequalities:

\begin{equation}
\left\{
\begin{aligned}
 &K_1b_3/a_3 > a_1b_3Y_{01}/a_3 + a_2b_3Y_{02}/a_3 + \tau\\
 &K_2 = b_1Y_{01} + b_2Y_{02} + \tau\\
 &K_3b_3/c_3 < c_1b_3Y_{01}/c_3 + c_2b_3Y_{02}/c_3 + \tau\\
 \end{aligned}
 \right.
 \label{Beq3}
\end{equation}

By solving Eq.(\ref{Beq3}), we obtain:
\begin{equation}
{Y}_{01} \le {Y}^+_{01}  = \frac{K_2(c_2a_3 - c_3a_2)+K_1(b_2c_3 - b_3c_2)+K_3(a_2b_3-a_3b_2)}{b_2(a_1c_3 - c_1a_3)+b_1(c_2a_3 - c_3a_2)+b_3(a_2c_1 - a_1c_2)}.
\label{B_Y01} 
\end{equation}
The upper bound of $q_{01}$ is:
\begin{equation}
q^+_{01} = a_0a_1Y^+_{01}.
\end{equation}

\textbf{Upper bound of ${q}_{02}$ and ${q}_{20}$ :}

We take ${q}^+_{02}$ as an example. According to Eq.(\ref{Eq_QY}), we have:
\begin{equation}
\begin{aligned}
&Q_{o\mu} \ge a_0Y_{00} + a_1Y_{01} + a_2Y_{02}; \\
&Q_{o\omega} \le c_0Y_{00} + c_1Y_{01} + c_2Y_{02} + (c_3 + c_4 +c_5 +....).
\end{aligned}
\end{equation}

Noting that $Q_{oo} = Y_{00}$ and defining $H_1 = Q_{o\mu} - a_0Q_{oo}$ and $L_3 = Q_{o\omega} - c_0Q_{oo} - (c_3 + c_4 + c_5 +....)$, we obtain:
\begin{equation}
\left\{
\begin{aligned}
&H_1 > Y_{01}a_1 + Y_{02}a_2\\
&L_3 < Y_{01}c_1 + Y_{02}c_2\\
 \end{aligned}
 \right.
 \label{Beq4}
\end{equation}

By solving Eq.(\ref{Beq4}), we obtain:
\begin{equation}
Y_{02} < Y^+_{02} =  \frac{H_1c_1 - L_3a_1}{a_2c_1 - a_1c_2}.
\end{equation}

The upper bound of $q_{02}$ is:
\begin{equation}
q^+_{02} = a_0a_2Y^+_{02}.
\end{equation}.
\textbf{Upper bound of ${q}_{11}$:} 

It's easily to calculate $\overline{q}_{11}$ by using $Q_{\mu\mu}$, $Q_{\mu o}$, $Q_{o \mu}$ and $Q_{o o}$:

\begin{equation}
\begin{aligned}
q_{11} < \sum^{\infty}_{n=1,m=1}a_na_mY_{nm} 
= Q_{\mu\mu} - a_0Q_{\mu o} - a_0Q_{o\mu} + (a_0)^2Q_{oo}.
 \end{aligned}
\label{Beq5}
\end{equation}\\

Thus the upper bound of $q_{11}$ is:
\begin{equation}
q^+_{11} = Q_{\mu\mu} - a_0Q_{\mu o} - a_0Q_{o\mu} + (a_0)^2Q_{oo}.
\end{equation}

\textbf{Lower bound of $q_{00} + q_{01} + q_{10} + q_{02} + q_{20} + q_{11}$:} 
Define ${q}_{sum} = \sum_{n=0}^2\sum_{m=0}^{2-m}q_{nm}$, $q_{t_1} = q_{00} + q_{01} + q_{10} + q_{02} + q_{20} $ and $q_{t_2} = q_{11}$. Obviously there is ${q}_{sum} = q_{t_1} + q_{t_2}$.

Firstly, we estimate the lower bound of $q_{t_1}$. It's obviously that:
\begin{equation} 
q_{t_1} > q^+_{t_1} =a_0[Q_{o\mu}+Q_{\mu o}-2(a_3 + a_4 + a_5 +....)]-(a_0)^2Q_{oo}.
\label{Beq6}
\end{equation}

Then we estimate the lower bound of $ q_{t_2}$. Similar to the estimation of $q^+_{01}$, by defining $T_1 = Q_{\mu\mu} - a_0(Q_{o\mu} + Q_{\mu o}) + (a_0)^2Q_{o o}$, $T_2 = Q_{\nu\nu} - a_0(Q_{o\nu} + Q_{\nu o}) + (a_0)^2Q_{oo}$ and $\tau' = \sum^\infty_{m=1}\sum^\infty_{n=1}b_mb_nY_{m,n} - (b_1)^2Y_{1,1}$, we obtain an equation set:

\begin{equation}
\left\{
\begin{aligned}
&\frac{T_1b_1b_2}{a_1a_2} < \frac{a_1b_1b_2Y_{1,1}}{a_2} + \tau' \\ 
&T_2 = (b_1)^2Y_{1,1} + \tau' \\
\end{aligned}
\right.
\label{Beq7}
\end{equation}

Solve the Eq.(\ref{Beq7}), we obtain:
\begin{equation}
Y_{11}\ge Y^-_{11}=\frac{T_1b_1b_2 - T_2a_1a_2}{a_1^2b_1b_2 - b_1^2a_1a_2},
\label{Beq8}
\end{equation}
and the lower bound of $q_{t_2}$ is:
\begin{equation}
q^-_{t_2} = a_1^2Y^-_{11}.
\end{equation}

Our target, i.e. the lower bound of $q_{sum}$ is:
\begin{equation} 
q^-_{sum} = q^-_{t_1} + q^-_{t_2}.
\end{equation}

\bibliography{sample}

\end{document}